\documentclass[]{aa}
\usepackage[varg]{txfonts}
\usepackage{graphicx}
\usepackage{fixltx2e}

\bibpunct{(}{)}{;}{a}{}{,} 

\begin{document}

\title{Numerical simulations of transverse oscillations in radiatively cooling coronal loops}

\author{N. Magyar\inst{\ref{inst1}\thanks{FWO (Fonds Wetenschappelijk Onderzoek) PhD fellow}}
\and T. Van Doorsselaere\inst{\ref{inst1}}
\and A. Marcu\inst{\ref{inst2}}}

\institute{Centre for mathematical Plasma Astrophysics, Department of Mathematics, KU Leuven, Celestijnenlaan 200B, bus 2400,
3001 Leuven, Belgium\label{inst1}
\and Babes-Bolyai University Cluj-Napoca, str. Mihail Kogalniceanu nr.1, Cluj-Napoca, Romania\label{inst2}}

\abstract {} 
{We aim to study the influence of radiative cooling on the standing kink oscillations of a coronal loop.} 
{Using the FLASH code, we solve the 3D ideal MHD equations. Our model consists of a straight, density enhanced and gravitationally stratified magnetic flux tube. We perturb the system initially, leading to a transverse oscillation of the structure, and follow its evolution for a number of periods. A realistic radiative cooling is implemented. Results are compared to available analytical theory.} 
{We find that in the linear regime (i.e. low amplitude perturbation and slow cooling) the obtained period and damping time are in a good agreement with theory. The cooling leads to an amplification of the oscillation amplitude. However, the difference between the cooling and non-cooling cases is small (around 6\% after 6 oscillations). In high amplitude runs with realistic cooling, instabilities deform the loop, leading to increased damping. In this case, the difference between cooling and non-cooling is still negligible, around 12\%. A set of simulations with higher density loops are also performed, to explore what happens when the cooling takes place in a very short time ($t_{\mathrm{cool}} \approx 100$ s). In this case, the difference in amplitude after nearly 3 oscillation periods for the low amplitude case is 21\% between cooling and non-cooling cases. We strengthen the results of previous analytical studies stating that the amplification due to cooling is ineffective, and its influence on the oscillation characteristics is small, at least for the cases shown here. Furthermore, the presence of a relatively strong damping in the high amplitude runs even in the fast cooling case indicates that it is unlikely that cooling could account alone for the observed, flare related undamped oscillations of coronal loops. These results may be significant in the field of coronal seismology, allowing its application to coronal loop oscillations with observed fading-out or cooling behaviour. } 
{}

\maketitle

\section{Introduction}

 Coronal loops gained much attention from the scientific community, both observationally and theoretically, since the first observational evidence of transverse MHD oscillations \citep{1999ApJ...520..880A,1999Sci...285..862N}, which had been theorized decades before their discovery \citep{1975...37..3,1976JETP...43..491R,1983SoPh...88..179E}.
 The study of coronal loop oscillations is important for two main reasons: on the one hand, standing modes are excellent tools for coronal seismology, a technique which determines coronal parameters that are hard to measure by other means. To improve the accuracy of this method, studies focused on increasingly realistic coronal loop models, including several effects that might have an influence on oscillation parameters,  e.g. loop curvature \citep[e.g.][]{2004A&A...424.1065V,2006ApJ...650L..91T} density stratification 
 \citep[e.g.][]{2005A&A...430.1109A,2005SoPh..229...79D}, variable cross section \citep[e.g.][]{2008A&A...486.1015V,2008ApJ...686..694R}, twisted magnetic field \citep[e.g.][]{2007SoPh..246..119R,2012A&A...548A.112T}. The transverse oscillations have been used
 for seismology to measure the magnetic field strength \citep{2001A&A...372L..53N,2007A&A...473..959V}, the density stratification \citep{2005ApJ...624L..57A}, the perpendicular structuring in the magnetic field \citep{2003ApJ...598.1375A} and the Alfv\'en travel time \citep{2007A&A...463..333A}. For a review on coronal seismology, see \citet{2012RSPTA.370.3193D}. On the other hand, since wave heating is a proposed mechanism for the mysteriously high temperature of the corona, ubiquitous propagating waves may contribute to energizing the coronal plasma (for a review on coronal heating, see, e.g. \citealt{2012RSPTA.370.3217P}). \par

 Since their first observation in 1998, numerous theoretical, numerical and observational works have been done (for a review on coronal loop oscillations, see \citealt{2009SSRv..149..199R}). In general, the coronal loops are not in a steady state and evolve during the oscillations. Previous studies mostly assume a static background loop. In a paper by \citet{2008ApJ...686L.127A} it was pointed out that the intensities of most observed coronal loops in a single EUV waveband vary, consistent with a plasma cooling scenario. They suggested that a proper MHD study of oscillating coronal loops should include the density and temperature changes due to the plasma cooling. In the first, zeroth-order analytical study of oscillating, radiatively cooling loops by \citet{2009ApJ...707..750M} it was shown that cooling leads to damping of the oscillations.
 However, in another study of the phenomenon \citep{2011A&A...534A..78R}, it was indicated that the cooling leads to an amplification of the oscillations. They found out that neglecting the flow caused by the radiative plasma cooling resulted in the damping behaviour in \citet{2009ApJ...707..750M}. 
 
 A common property of observed transverse coronal loop oscillations is that they are damped quickly, usually within a few oscillation periods. It is now generally accepted that the main damping mechanism is resonant absorption \citep{1991SoPh..133..227S,1992SoPh..138..233G,2002ApJ...577..475R,2002A&A...394L..39G}, transferring energy from the global kink mode to local azimuthal Alfvén modes in the boundary layer of the loop structure, where the two frequencies match (for a review on theoretical results, see \citealt{2011SSRv..158..289G}). Less frequently, nearly undamped or even growing transverse oscillations are observed \citep{2012ApJ...751L..27W}.  Decayless low amplitude kink oscillations, present in loops before and after high-amplitude flare triggered damped kink oscillations can be explained in terms of a damped linear oscillator excited by a continuous low amplitude harmonic driver \citep{2013A&A...552A..57N,2013A&A...560A.107A}. Some examples of observed undamped high amplitude coronal loop oscillations can be found in \citet{2002SoPh..206...99A} and \citet{2011ApJ...736..102A}. In the latter paper, it was concluded that this could happen only if the thickness of the boundary layer (a radially inhomogeneous outer layer of the coronal loops) is much smaller than the loop radius, in order to minimize the damping due to resonant absorption. It was also evident from the observations that the loop was cooling during the particular undamped oscillation event. It seemed a rather natural explanation that the amplification due to cooling may counterbalance the damping due to resonant absorption, explaining the undamped oscillations, as it was firstly suggested in \citet{2011SoPh..271...41R}. The most elaborate analytical study regarding time-dependent kink oscillations \citep{2011A&A...534A..78R} considered the simultaneous effects of both damping due to resonant absorption and amplification due to plasma cooling. The conclusion was that, for typical boundary layer thicknesses, the amplification due to plasma cooling can account for the observed undamped oscillations only if the cooling happens quickly, on the order of the oscillation period.
 However, in the paper, the effects of cooling and resonant damping have been studied under the assumption that both the characteristic cooling time and damping time are much larger than the characteristic oscillation period, using the WKB method. Thus, it is questionable whether the derived equations remain valid for rapid cooling, i.e. for cooling times on the order of oscillation periods. On the other hand, neglected nonlinear behavior may considerably change the outcome, for example the Kelvin-Helmholtz instability at the tube boundary \citep{1983A&A...117..220H,1994GeoRL..21.2259O,2008ApJ...687L.115T}, and for high amplitudes, the ponderomotive forces \citep{2004ApJ...610..523T}. 
 The aim of this study is to further investigate the effects of radiative plasma cooling on the fundamental standing kink oscillation, by means of numerical analysis.

\section{Numerical Model}

 The 3D numerical model consists of a straight, density enhanced magnetic flux tube (also referred to as loop in what follows) embedded in a background plasma. We aim to model a coronal active region loop. Initially, the system is permeated by a uniform magnetic field directed along the flux tube, i.e. in the $z$ direction. Gravity is included, and it varies along the flux tube sinusoidally, i.e. it has zero value at the loop center ($z = 0$) and maximum/minimum values at footpoints ($z = \pm L/2$), in order to mimic the component of gravity parallel to the magnetic field in a semi-circular loop. Thus, we have stratification along the loop according to the hydrostatic equilibrium both inside and outside the loop (Fig. 1):
  
 \begin{equation}
     \frac{dp_{\mathrm{i,e}}(z)}{dz} = - \rho_{\mathrm{i,e}}(z)\ g\ \sin\left(\frac{\pi}{4 L}z\right)
 \end{equation}
 where $p$ is thermal pressure, $\rho$ is mass density, the subscripts i,e denote the interior and exterior plasma respectively, and $g = 274$ m/s$^2$ is the Sun's surface gravity. 
 
 \begin{figure}
    \centering
    	\includegraphics[width=0.5\textwidth]{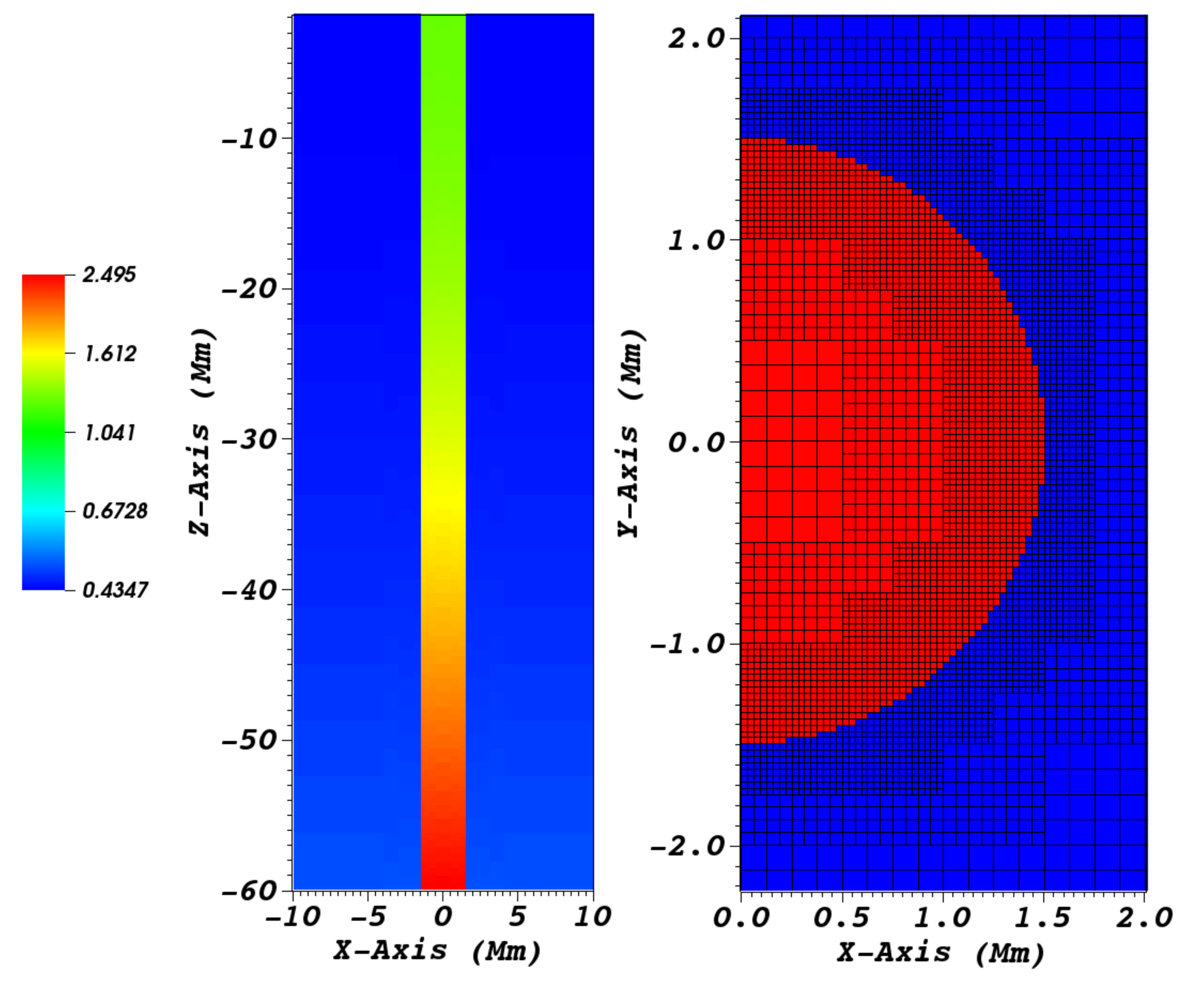}  
    	\caption{Plots showing the mass density at $t = 0$ : cross section along the axis of the loop (left) and perpendicularly to the loop at its footpoint (right). In the right plot, the mesh (numerical cells) is also shown.  }
    	\label{equil}
 \end{figure}

Due to stratification, there will be a pressure imbalance at the loop boundary, which leads to a jump in total pressure. This is rapidly equilibrated once the simulation is started, resulting in a slightly increased and stratified magnetic field inside the loop. The perturbation caused by this imbalance has no effect on the long-timescale evolution, even if we do not let the system settle before we trigger the kink mode. We implement a step density profile for the loop. However, due to the numerical diffusivity, a thin boundary layer evolves at the interface. The presence of this inhomogeneous layer allows for resonant absorption in the system. Values of the principal physical parameters used in the simulation are given in Table ~\ref{table}.
 
  \begin{table}
  \caption{\label{t7}The values of principal physical parameters used in the simulations.}
  	\centering
  	\begin{tabular}{lccc}
  		\hline\hline
  		Parameter & Value \\ 
  		\hline
  		Loop length ($L$)& 120 Mm \\
  		Loop radius ($R$)& 1.5 Mm \\
  		Magnetic Field & 12.5 Gauss \\
  		Loop footpoint density ($\rho_\mathrm{fi}$) & $2.5 \cdot 10^{-12}$ kg/m$^3$ \\
  		Density ratio at footpoint ($\rho_\mathrm{fi}/ \rho_\mathrm{fe}$)& 5 \\
  		Loop temperature & 0.9 MK \\
  		Background plasma temperature & 4.5 MK \\
  		Plasma $\beta$ & 0.06 \\
  		\hline
  	\end{tabular}
  	\label{table}
  \end{table}
 \subsection{Radiative Cooling}
 
 At low coronal temperatures (below 1 MK), the radiative cooling time is of the order $t_{\mathrm{rad}} \approx 10^3$ s for typical coronal loops. This is 2 orders of magnitude smaller than the conductive cooling time, making the radiative loss the dominant cooling mechanism \citep{2008ApJ...686L.127A}. Thus, we neglect thermal conduction. Potential implications of this are discussed in the Conclusions section.
 We do not consider any heating mechanisms present during the simulation period. \par
\begin{figure}
  	\centering
  	\includegraphics[width=0.5\textwidth]{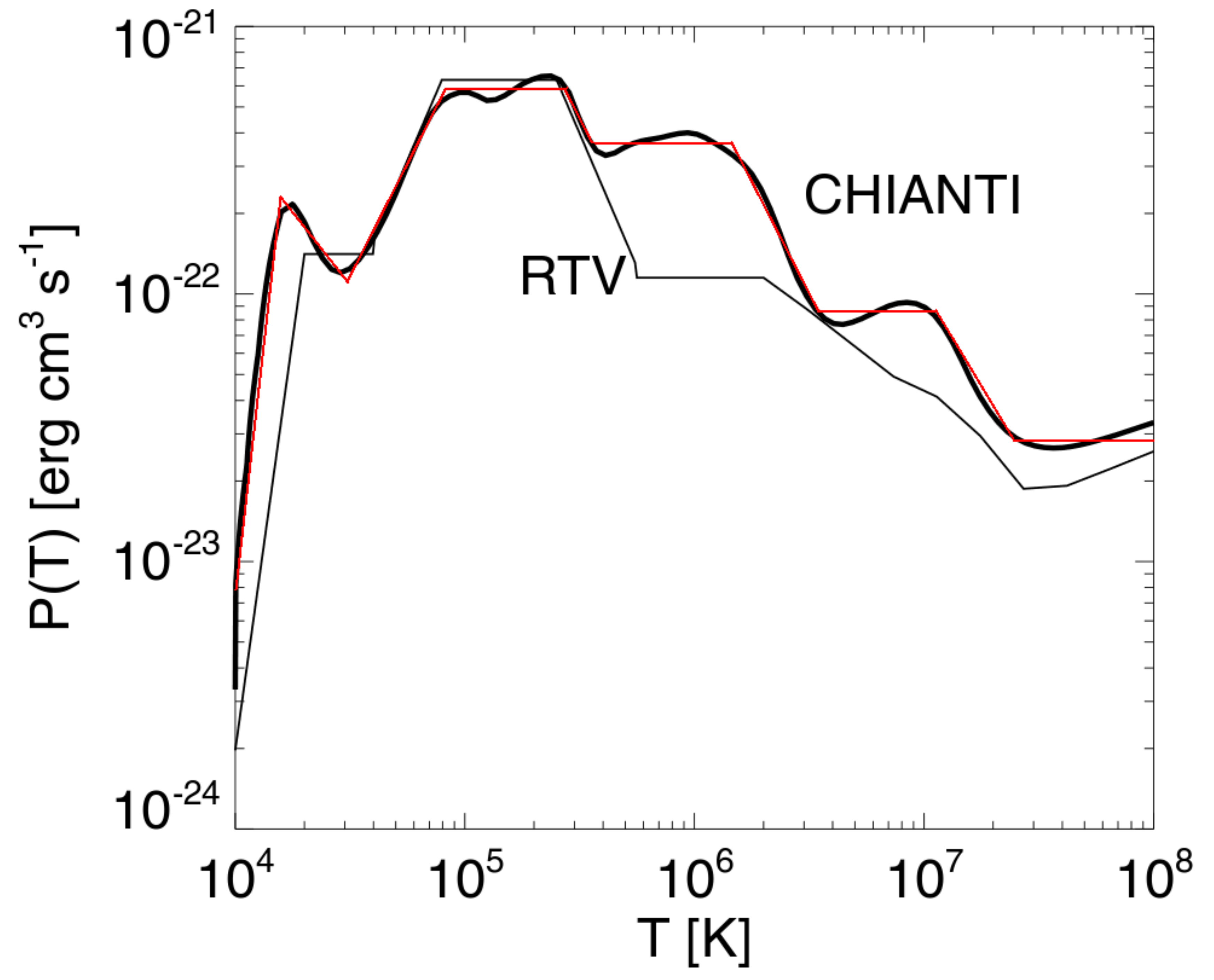}
  	\caption{Plasma emissivity as a function of temperature, according to \citet{1978ApJ...220..643R} (thin solid line), and to version 7 of the CHIANTI spectral code \citep{2012ApJ...744...99L} (thick solid line). The red line represents the piecewise emissivity used in our simulations. Adapted from \citet{2012A&A...543A..90R}.}.
  	\label{cool}
  \end{figure}

 
 The cooling module used in the simulation models the radiative loss rate $E(T)$ for an optically thin plasma \citep{1982ApJ...252..791P}:
 \begin{equation}
 E(T) = P(T)n_\mathrm{i} n_\mathrm{e}
 \end{equation}
 where $n_\mathrm{i}$ and $n_\mathrm{e}$ is the ion and electron number density, which are considered to be equal, and $P(T)$ is the \textit{plasma emissivity function}, which is strongly dependent on the temperature. In the simulations, the radiative loss rate $E(T)$ is calculated at each time step and is used then in the energy equation:
 \begin{equation}
	 \frac{\partial \rho E}{\partial t} + \nabla \cdot (\textbf{v}(\rho E + p_T)-\textbf{B}(\textbf{v} \cdot \textbf{B})) = \rho \textbf{g}\cdot \textbf{v} + E(T)
 \end{equation}
 where $E = \frac{1}{2}v^2 + \epsilon + \frac{1}{2}\frac{B^2}{\rho}$ is the specific total energy with $\epsilon$ the specific internal energy, $p_T = p + \frac{B^2}{2}$ the total pressure and the other variables have their usual meaning. The radiative cooling affects the whole system, however the background plasma cools much slower than our loop structure, due to its higher temperature and lower density.
 The used function for $P(T)$  follows closely the one computed with CHIANTI \citep{2012ApJ...744...99L}, shown in Fig.~\ref{cool}, where it can also be seen that for some temperature ranges (e.g. 0.5-2 MK), the emissivity calculated with CHIANTI is four times larger than previously assumed, thus cooling the coronal loops faster than with older emissivity functions. This is due to improvements in atomic models with the inclusion of more accurate atomic data and transition rates, or lines previously unavailable. 
 
  For comparison with analytical results, we use a decreased cooling (realistic cooling multiplied by a constant $\textless 1$ factor), in order to allow for a slower and linear plasma evolution. 
 
 \subsection{Perturbation and boundary conditions} 
 \label{sec:bc}
  
 Initially, we perturb the transverse component of velocity inside the loop with a pulse $v_{0_y} = A_\mathrm{v} v_\mathrm{A,fi} \cos (\pi z/L)$, which excites a standing kink mode.  Here, $v_\mathrm{A,fi} \approx 0.7$ Mm/s is the Alfvén velocity inside the loop, at its footpoints, and $A_v$ is the dimensionless, relative amplitude of the perturbation ($A_\mathrm{v} \ll 1$ for linear regime). We focus on modeling standing kink coronal loop oscillations triggered by a flaring event, and not the recently observed ubiquitous small amplitude kink oscillations \citep{2013A&A...552A..57N,2013A&A...560A.107A}, which are triggered by footpoint excitations.The initial perturbation acts only inside the loop, and is constant in radial direction. Note that, due to stratification, the pulse does not correspond exactly to the fundamental kink eigenmode, but higher harmonics are also excited to a small extent (see \citealt{2005A&A...430.1109A}).
 Exploiting the symmetric properties of standing kink waves, only half of the loop in both longitudinal ($z$ axis) and transverse ($x$ axis) direction is modeled, thus reducing the computational time four-fold. For these planes, in order to simulate the whole loop, symmetric boundary conditions are used, which are the following: in the $x-y$ plane (at the apex), $v_z, B_y, B_x$ change sign, thus are antisymmetric, while the other variables are symmetric. In the $y-z$ plane (the plane cutting the loop in half along it), only $v_x$ and $B_x$ are antisymmetric. \\
 Note that we do not employ a realistic solar atmosphere model (i.e. with photosphere, chromosphere, transition region). In order to mimic coronal loops anchored in the dense photosphere, the loop footpoint is fixed. At this boundary, the line-tying condition is used, setting the velocities in all directions to zero, implying that plasma can not leave the domain through loop footpoints.  In order to validate our use of the line-tying boundary condition in getting the right behaviour of the downflow at the loop footpoints, we ran simulations with a realistic atmosphere, without transverse velocity perturbations. From the results of these simulations we can state that plasma is not evacuated from the loop structure, but rather accumulates at loop footpoints initially. Furthermore, the thermodynamic properties and the flow inside the loop are not significantly altered by the line-tying condition (this can be appreciated by comparing Fig.~\ref{amp2-dens} and Fig.~\ref{amp3-dens}). The correct condition for the other variables at this boundary is that of a continuation of the hydrostatic equilibrium with constant temperature in the ghost cells. The other boundaries, in order to minimize their influence on the dynamics, are placed at a safe distance from the loop ($13\ R$ in the direction of the displacement, i.e. $y$ axis and $4\ R$ for the $x$ axis). At these boundaries, the \textit{outflow} or \textit{open} boundary condition is used, which allows waves to leave the domain. 
 \par
 
 \subsection{Numerical method and grid}
 
  The 3D ideal MHD problem is solved using the FLASH code, which implements a second order unsplit Godunov method \citep{2009JCoPh.228..952L,2013JCoPh.243..269L} and constrained transport for keeping the solenoidal constraint on the magnetic field. We use the `mc' slope limiter and the Roe-type solver. An adaptively refined mesh is used, in order to have high resolution only in the domain of interest, i.e. around the loop, with 5 levels of refinement. The variable used for triggering mesh refinement (calculated with L{\"o}hner's error estimator) is the density. Initially, the grid consists of $24 \times 40 \times 32$ numerical cells, thus the resolution is bigger in the $x-y$ plane, in order to resolve the small-scale phenomena which appear around the loop edge, such as instabilities and resonant absorption. In the $z$ direction, the solution is smooth (wavelength of the order of box length). The effective resolution then (if the whole box would be refined), with 5 levels of refinement is $1280 \times 384$ in the $x-y$ plane, which translates in cell sizes of 31.25 km, or 0.02 $R$. Test simulations with 6 levels of refinement show an effect on the small scales (instabilites) present in the perpendicular direction, but no relevant changes to oscillation characteristics (period, damping rate).
  
 \section{Results and discussion}
 
 \subsection{Low amplitude perturbations}
 
  \begin{figure*}
  	\centering
  	\includegraphics[width=0.75\textwidth]{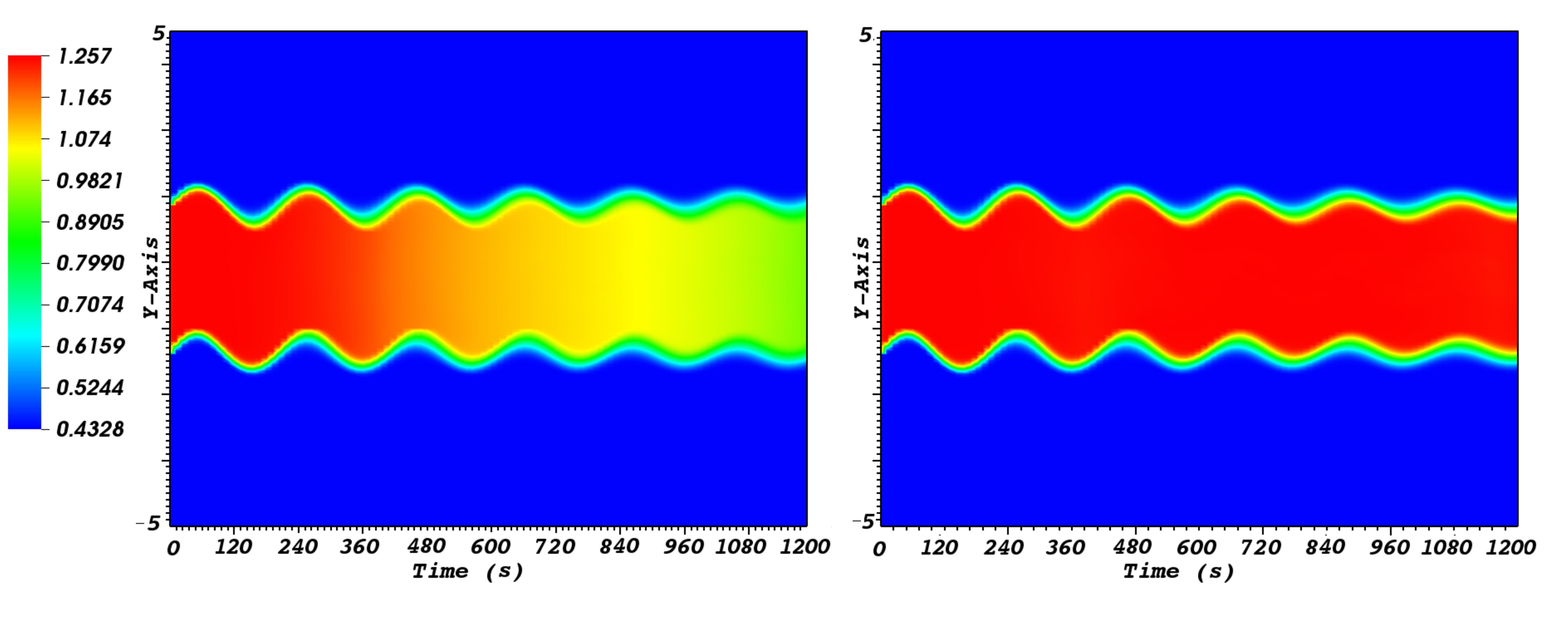}
  	\caption{Time-slice plot of the density at the loop apex ($z = 0.0$), for cooling (left) and non-cooling (right) cases, showing the evolution of the oscillation over time. The colour scale shows the density and is given in units of $10^{-12}$ kg/m$^3$. The colour scale is common for the two images.}
  	\label{amp-evol}
  \end{figure*}
  
  \begin{figure*}
  	\centering
  	\includegraphics[width=0.75\textwidth]{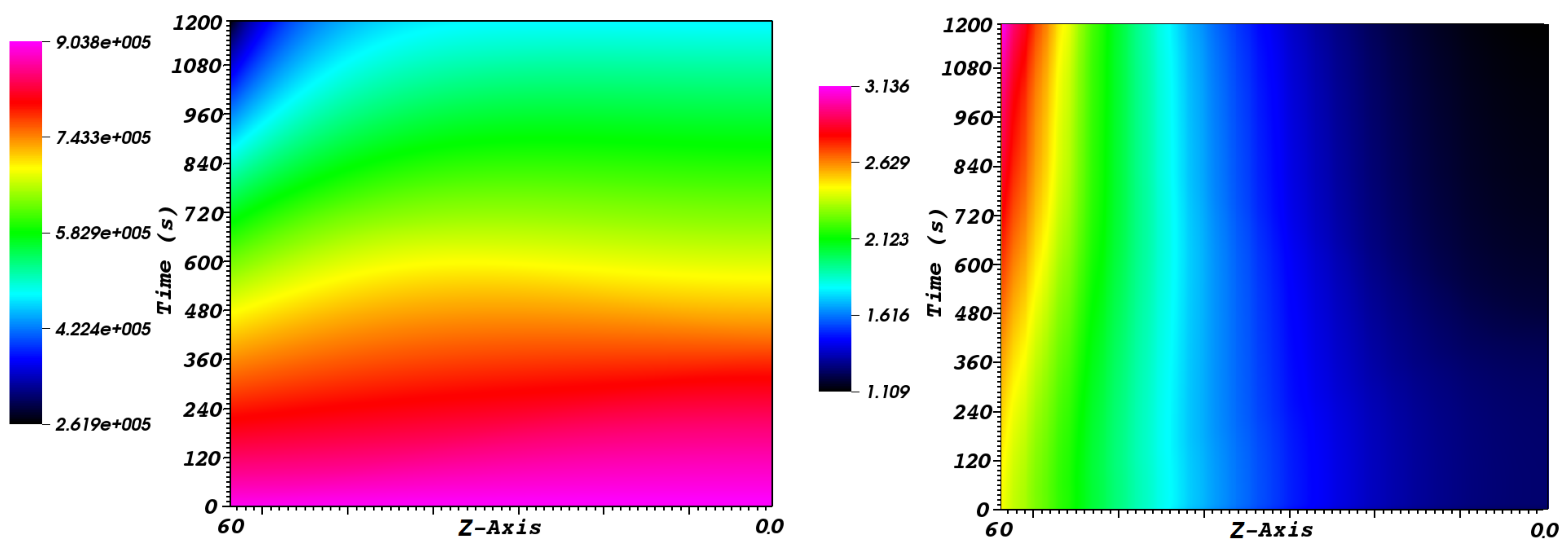}
  	\caption{Time-slice plot of the evolution of both the temperature (left) and density (right) at the axis of the loop, over time, for the cooling case. Temperatures are in K, while density is in units of 10$^{-12}$ kg/m$^3$. The horizontal $z$ axis spans from the loop footpoint (left end) to the apex (right end).}
  	\label{amp-dens}
  \end{figure*}
 
 Initially, we consider relatively low amplitude initial velocity perturbations ($A_\mathrm{v} = 0.02$ or $v_{0_y} = 14$ km/s), in order to remain in the linear regime, necessary for comparing our results with available analytical calculations. This also implies that we change from the realistic radiative cooling to a slower cooling, by multiplying the radiative loss function with a constant factor to get a cooling time $t_{\mathrm{cool}} \approx 1500$ s. Simulations are performed both with and without cooling. The perturbation leads to a maximum displacement of the loop at the apex of 0.15 R or 225 km, thus a peak-to-peak displacement of 450 km. 
 We let the system evolve until $t_\mathrm{f} = 1200$ s, in which we observe approximately 6 periods of oscillation, with a mean period of 204 s for the non-cooling case, very close to the analytically predicted value for the standing fundamental kink oscillations of a uniform flux tube with densities (inside and outside) equal to the weighted mean density of our stratified loop \citep{1983SoPh...88..179E,2005A&A...430.1109A}:
 \begin{equation}
 P = \frac{2L}{C_\mathrm{k}} = 2L\sqrt{\frac{\langle{\rho_\mathrm{i}}\rangle+\langle{\rho_\mathrm{e}}\rangle}
 	 {\langle{\rho_\mathrm{i}}\rangle\langle{v_\mathrm{A,i}}\rangle+\langle{\rho_\mathrm{e}}\rangle\langle{v_\mathrm{A,e}}\rangle}} = 206\ \mathrm{s}
 \end{equation}
 where $v_\mathrm{A,i,e}$ are the internal and external Alfvén speeds, and the weighting function used to obtain the mean values is $\cos^2\left(\pi\frac{z}{L}\right)$. The weighting function represents the wave energy density distribution along the loop of the fundamental mode (see \citealt{2005A&A...430.1109A}).
 In the non-cooling case, the oscillation is damped due to the energy transfer between the global kink mode and local torsional Alfvén modes, i.e. resonant absorption, resulting in a damping time $\tau_\mathrm{D} \approx 1074$ s. If we assume that the density varies sinusoidally in the inhomogeneous layer (which is not exactly the case for our data) the theory predics a damping time    \citep{2002ApJ...577..475R,2002A&A...394L..39G,2005A&A...441..361A}: 
 \begin{equation}
 \tau_\mathrm{D} = \frac{2}{\pi}\frac{a}{l}\left(\frac{\langle{\rho_\mathrm{i}}\rangle+\langle{\rho_\mathrm{e}}\rangle}
 {\langle{\rho_\mathrm{i}}\rangle-\langle{\rho_\mathrm{e}}\rangle}\right)P = 1463\ \mathrm{s}
 \end{equation}
 where $\frac{a}{l}$ is the total to inhomogeneous layer width ratio.
 Keeping in mind the uncertainties of the inhomogeneous layer profile and width (for a linear profile with the same width, $\tau_\mathrm{D} = 927$ s), we can just state that the damping time obtained from the simulation lies inside the range of values predicted by the theory (see \citealt{2013ApJ...777..158S,2014ApJ...781..111S}).
 Now we look at the differences between the two runs, i.e. oscillations in cooling and non-cooling case. The obvious difference between the two lies in the longitudinal evolution. More specifically, cooling of the plasma induces a flow inside the loop, which rearranges plasma towards the footpoints. Thus, there is a density increase close to the footpoints and a decrease at the loop tops (see Fig.~\ref{amp-evol} and Fig.~\ref{amp-dens}). 
 
\begin{figure}
   	\centering
   	\includegraphics[width=0.5\textwidth]{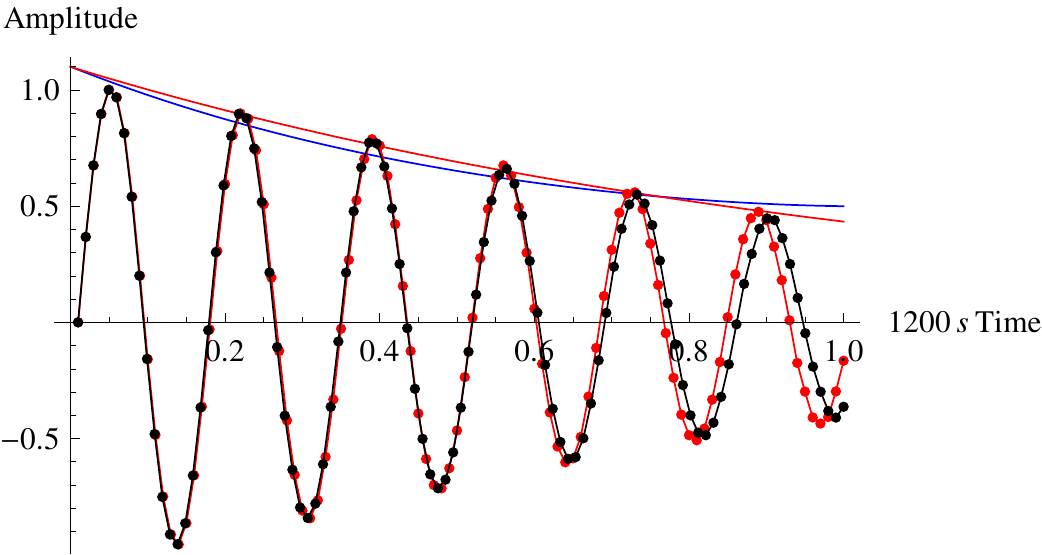}  
   	\caption{Graph of normalized loop displacement at the apex for both cooling (red dots) and non-cooling (black dots), over time. The red curve represents the best-fit exponential decay, while the blue curve is the analytically predicted amplitude (\citealt{2011A&A...534A..78R}), both shown for the cooling case. The displacements were obtained by center-of-mass tracking in the apex cross-section of the loop. }
   	\label{amp}
   \end{figure}

 This effect can be easily seen from the continuity equation: if we have a time-dependent density (due to cooling), it will give rise to a varying velocity. The resulting flow speeds in the low amplitude run are of the order of few tens of km/s , being at the lower boundary of the observed downflow speeds, in the range $40 - 120$ km/s \citep{2001SoPh..198..325S}.
 As mentioned in Section~\ref{sec:bc} there are no outflows throughout the loop footpoints. Simulations with included realistic atmosphere (hyperbolic tangent temperature, see, e.g. \citealt{2010A&A...521A..34K}) without velocity perturbations show that plasma evolution near the loop footpoint is approximated well by the simpler line-tying boundary condition (see again Fig.~\ref{amp2-dens} and Fig.~\ref{amp3-dens}). We could not employ the realistic atmosphere in our study of the oscillations because of disturbances originating from the transition region due to radiative cooling, which could have altered the oscillation characteristics. Now, we compare our oscillation amplitude evolution with the analytically predicted one from \citet{2011A&A...534A..78R}, which includes the effects of both resonant absorption and cooling on the oscillation characteristics. We solve Eq. (98) from the paper numerically for our parameters and obtain the amplitude over time, resulting in a damping time $\tau_\mathrm{D} \approx 1090$ s. Note that in the analytical studies it is assumed that the hydrostatic formula is valid throughout the evolution, arguing that the flow effect on the density distribution in weak. This implies that the loop footpoint density is considered constant and that there is a net outflow of plasma through these footpoints.  Thus, the comparison with analytical results should only be qualitative. However, the two evolutions (numerical and analytical) are in a good agreement (Fig.~\ref{amp}).
 
  We can see that the difference in amplitude between the cooling and non-cooling cases is minimal ($\approx 6\%$), thus the effects of cooling on the oscillation are very small. The efficiency of amplification due to cooling strongly depends on the characteristics of the loop (boundary layer thickness ($l$), density scale height, etc.), but probably most importantly on its hydrodynamic evolution, reflected in the cooling time. In our case, the inefficiency might come from the relatively thick boundary layer ($l/R \approx 0.19$), created by numerical diffusion. In \citet{2011A&A...534A..78R}, it is stated that, for typical conditions and cooling times, for the oscillations of coronal loops to be undamped, the boundary layer should be extremely thin ($l/R \approx 0.02$). However, such a thin boundary layer might be very unlikely for oscillating solar coronal loops, as will be argued in what follows.
  
   \begin{figure*}
   	\centering
   	\includegraphics[width=1.0\textwidth]{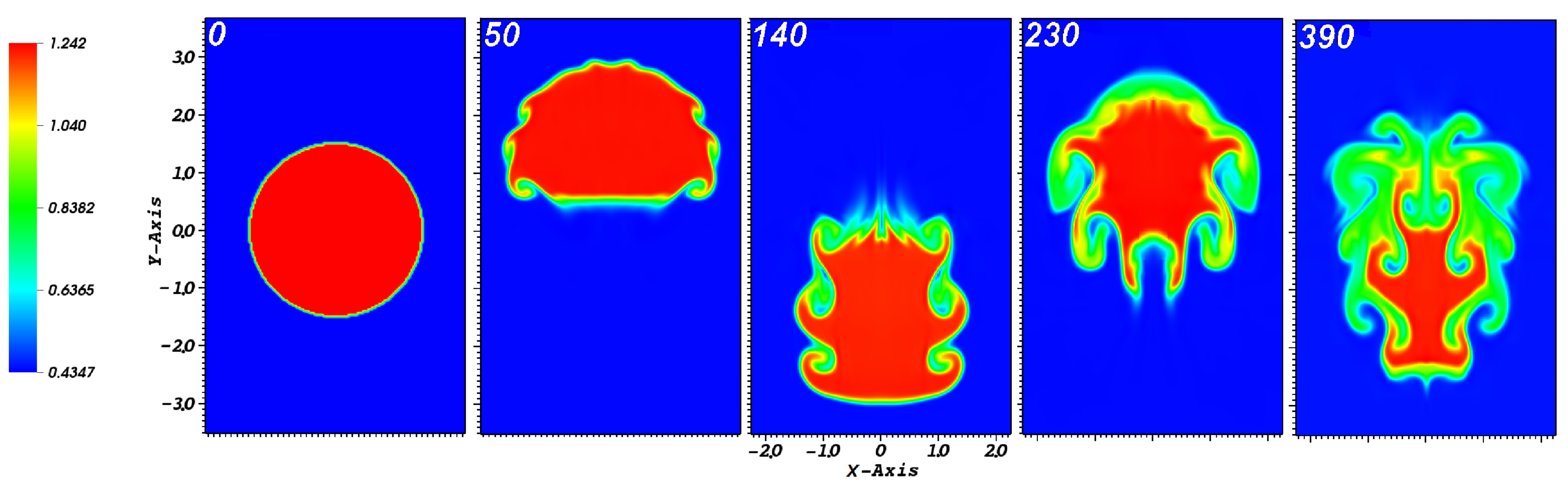}  
   	\caption{A sequence of cross-section plots of the density at the loop apex, at different times (written at the left-top of each plot, in seconds), for the high amplitude case. Axis units are in Mm, while density is in units of 10$^{-12}$ kg/m$^3$}
   	\label{instab}
   \end{figure*}

  \subsection{High amplitude perturbations}
  
\begin{figure}
     \centering
     	\includegraphics[width=0.5\textwidth]{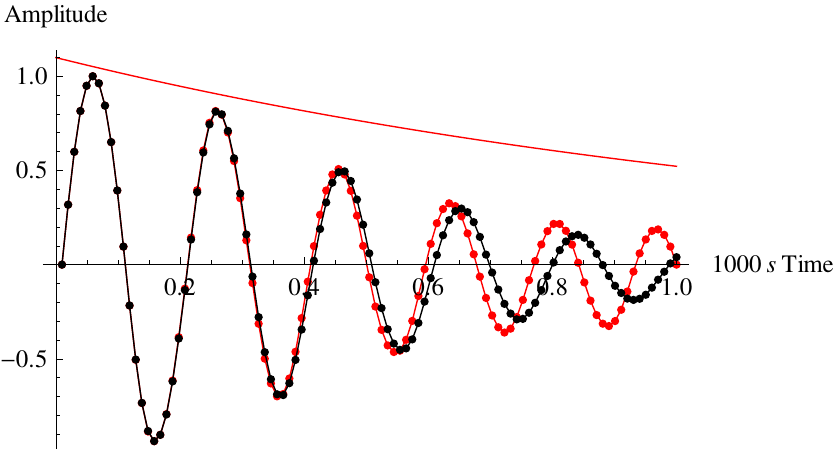}  
     	\caption{Same as in Fig.~\ref{amp} ( cooling with red dots and non-cooling with black dots), but for the high amplitude case. The red curve is the best-fit exponential decay for the non-cooling, low amplitude case (thus showing the added damping due to the presence of the instability). Note that $t_\mathrm{f} = 1000$ s and a realistic cooling for the high-amplitude runs, with $t_{\mathrm{cool}} \approx 800$ s.}
     	\label{amp2}
    \end{figure}

   Now, we consider a higher initial perturbation, with  $A_v = 0.1$, thus a velocity perturbation of $v_{0_y} = 70$ km/s (5 times bigger than in the low amplitude setup), which leads to an initial displacement of $1.6$ Mm, or around one loop radius. The displacement produced by the high perturbation is, however, at the lower boundary of the flare related, typically observed displacements (see \citealt{2011ApJ...736..102A}, \citealt{2008ApJ...687L.115T}). The realistic radiative loss used now leads to a faster cooling ($t_{\mathrm{cool}} \approx 800$ s), and new features are observed when compared to the previous linear evolution, the most important for the oscillation characteristics being the presence of instabilities at the loop edge, namely the Kelvin-Helmholtz instability, which deforms the loop drastically (see Fig.~\ref{instab}). In \citet{2014ApJ...787L..22A} it is stated that even for low amplitudes, the instability sets in. However, in our low amplitude case, the shear instabilities does not evolve, or their growth time is longer than our 6 period simulation time. This might be caused by the higher radius-length ratio of our loop or (and) a higher numerical viscosity of the scheme that we use, which might greatly affect the growth rate of the Kelvin-Helmholtz instability. \\
   From Fig.~\ref{amp2} we see that the damping of high amplitude oscillation is faster than in the low amplitude case. This is an important effect of the instabilities present in the system. The damping is faster for two reasons: firstly, the development of the instability dissipates kinetic energy, and secondly, due to the mixing caused by the instability, a wider inhomogeneous layer develops around the loop, which affects the effectiveness of resonant absorption. \par
    
   \begin{figure*}
   	\centering
   	\includegraphics[width=0.75\textwidth]{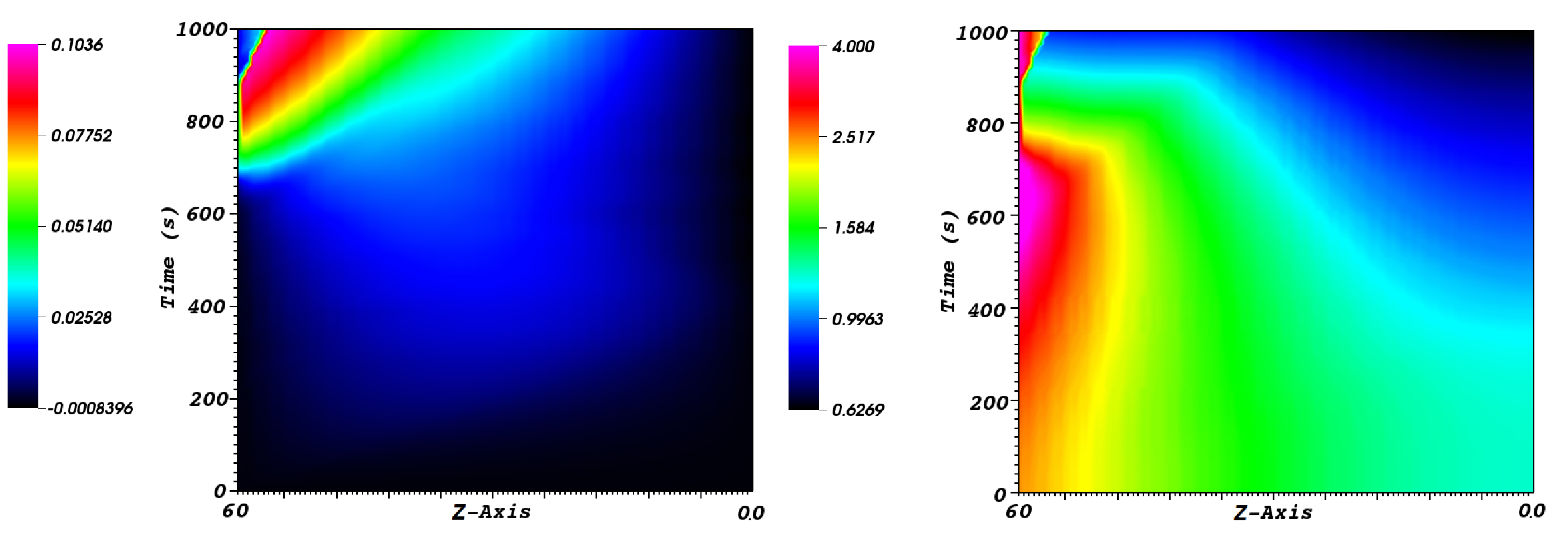}  
   	\caption{Time-slice plots showing the evolution of longitudinal flow speed ($v_z$, left) and the density (right) for the cooling case in the high amplitude run, along the loop axis, over time. Flow speed is in Mm/s, while density in units of 10$^{-12}$ kg/m$^3$.}
   	\label{amp2-dens}
   \end{figure*}    

   Compared to Fig.~\ref{amp}, the effects of cooling on the wave amplitude become measurable after $t \approx t_{\mathrm{cool}}$, but are not significant when looking at the whole evolution.
   The best-fit exponential damping time is 805 s for the cooling and 710 s for the non-cooling loop. Thus, the cooling only results in a 12 \% weaker damping. 
   Another effect is the decrease of the oscillation period for the cooling case, an effect which was shown in analytical studies \citep{2009ApJ...707..750M,2011SoPh..271...41R}. For our cooling case, the ratio between the initial oscillation period and that at the last anti-node (at $\mathrm{t_f} \approx 1000$ s, with the periods measured by hand) is $P_{\mathrm{t_{f}}}/P_i \approx 0.85$, which is a smaller deviation than predicted by the linear theory after the same amount of time ($P_{\mathrm{t_{f}}}/P_i \approx 0.57$, by solving Eq. (28) from \citealt{2009ApJ...707..750M} with our parameters). This deviation comes from the already mentioned differences between the analytical and numerical studies, i.e. in the analytical study, the footpoint density is kept constant thus plasma leaves the loop while in our studies, it accumulates at the footpoint.  \\
   Another feature present in the high amplitude runs (due to the realistic radiative cooling) is the different late stage evolution (compare the density evolution from Fig.~\ref{amp-dens} to that of Fig.~\ref{amp2-dens}). At around $t = t_{\mathrm{cool}} \approx 750$ s, there is a sudden draining of mass towards the footpoints with flow speeds of up to 100 km/s, (downflow speeds typically observed in the corona), generated by a runaway cooling of the accumulated plasma. However, this effect is of secondary importance for the present study. \\
   As stated above, the displacement observed in the high amplitude setup is still small compared to the typical displacements observed for flare related oscillating coronal loops. Thus, if the instabilities truly develop in oscillating solar coronal loops, for which there is no observational evidence yet (\citealt{2008ApJ...687L.115T}, but see \citealt{2014ApJ...787L..22A} where they claim it could be observed as loop strands), the existence of a very thin inhomogeneous layer for several oscillation periods is highly unlikely, thus implying heavy limitations on the effectiveness of cooling induced amplification.

\subsection{High density runs}

\begin{figure}
     \centering
     	\includegraphics[width=0.5\textwidth]{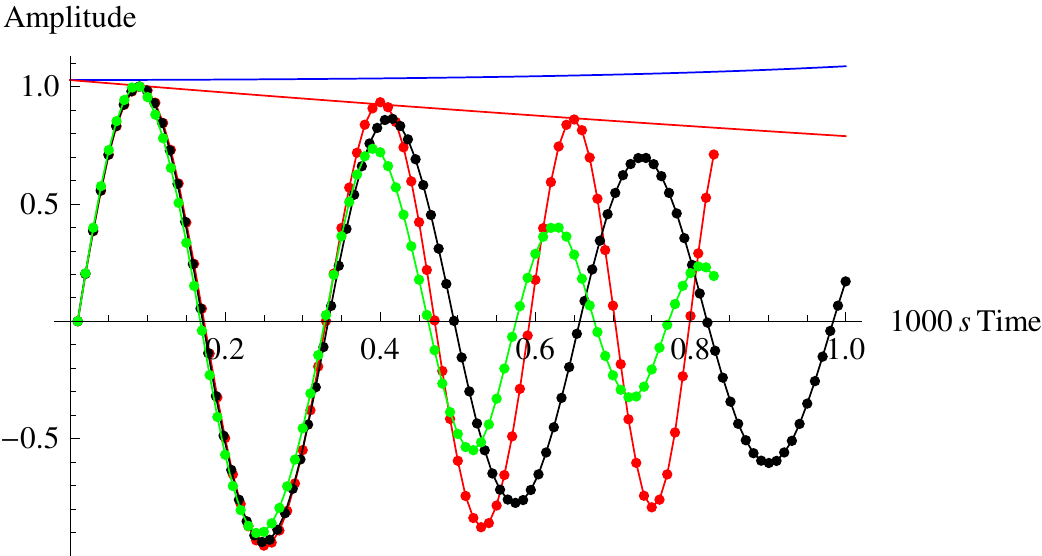}  
     	\caption{Normalized displacement amplitudes at the apex over time, for 3 cases:
     	low amplitude perturbation with cooling (red dots), non-cooling (black dots) and high amplitude perturbation with cooling (green dots). The red curve represents the best-fix exponential decay, while the blue curve is the analytically predicted displacement, for the low amplitude with cooling case. The displacements were obtained by center-of-mass tracking in the apex cross-section of the loop.}
     	\label{amp3}
    \end{figure}

\begin{figure*}
     \centering
     	\includegraphics[width=0.75\textwidth]{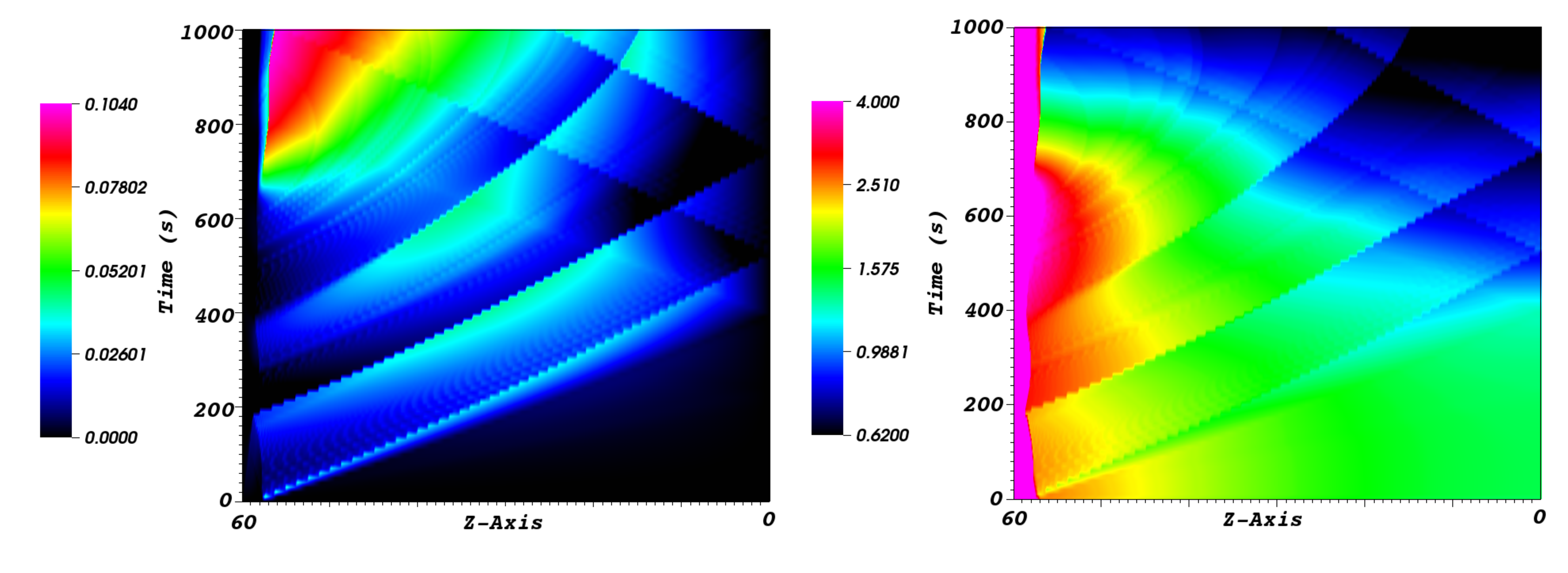}  
     	\caption{Time-slice plots showing the evolution of flow speed ($v_z$, left) and the density (right) with included lower solar atmosphere, for the same conditions as in Fig.~\ref{amp2-dens}, along the loop axis, over time. The saw-tooth appearance of wave fronts is due to the limited number of snapshots (100). Flow speed is in Mm/s, while density in units of 10$^{-12}$ kg/m$^3$. }
     	\label{amp3-dens}
    \end{figure*}

To extend the scope of our study and conclusions, a series of simulations with fast cooling were run. The faster cooling was achieved by setting the density ratio three times higher than in the previous setups ($\rho_\mathrm{fi}/ \rho_\mathrm{fe}$ = 15) while keeping the same temperature and magnetic field strength, resulting in a footpoint density inside the loop of $\rho_\mathrm{fi} = 7.5 \cdot10^{-12}$ kg/m$^3$. Note that this results in an increased plasma $\beta$ inside the loop. The cooling time is extremely short for these runs ($t_{\mathrm{cool}} \approx 100$ s), cooling the near-MK plasma in the loop to chromospheric temperatures within 200 s. Although it is much faster than the usually observed cooling times in the range 500-2000 s (see, e.g., \citealt{2008ApJ...686L.127A}), it is important to see whether such a high energy loss can alter significantly the oscillation characteristics. The resulting flow towards the footpoint is steadily increasing, peaking at 140 km/s around $\mathrm{t_f} \approx 830$ s, the end of simulation time.
The resulting displacements over time for three cases, cooling with low and high amplitudes (perturbations with the same fraction of footpoint Alfv\'en speeds as in the previous runs), and non-cooling with low amplitude, can be seen in Fig.~\ref{amp3}. \\
For the low amplitude case, the linear theory \citep{2011A&A...534A..78R} predicts that the oscillation amplitude should grow in time, in discrepancy with our results which show damping behaviour. However, the effects of the cooling in the low amplitude case are now stronger than for our previous runs, as expected: after $\approx$ 750 s, or 3 maximum displacements, the cooling case has a 21$\%$ higher amplitude than the non-cooling case.
The effect on the oscillation period is also more significant, the ratio of oscillation periods between the two cases being $P_{\mathrm{cool}}/P_{\mathrm{nocool}} \approx 0.6$ after the same time. \\
Looking at the high amplitude run, we still observe a strong damping, despite the fast cooling. This indicates that observed undamped high amplitude kink oscillations of coronal loops are likely not due solely to plasma cooling.

\section{Conclusions}

   We aimed to perform the first three dimensional numerical study of a particular and often observed phenomenon: coronal loop oscillations in a cooling coronal loop. For a better estimate of hard-to-measure parameters using coronal seismology, theoretical models must take into account several physical effects that might have an influence on observable oscillation characteristics, and cooling is one of them. We find that, in the linear regime (i.e. small amplitude oscillations and long cooling times), the effect of cooling is negligible. This may be attributable to the relatively thick inhomogeneous layer in our simulations, which arises solely due to numerical diffusion. Even if there are differences regarding boundary conditions between the available analytical results and our simulations, the resulting amplitude evolutions are in a good agreement.  \par 

   Increasing the initial velocity perturbation five-fold, resulting in a total displacement which lies at the low end of the observed, flare related kink oscillations and employing realistic radiative losses shows different evolution compared to the linear regime run: instabilities strongly affect the outer layers of the loop, and mixing causes a wider inhomogeneous layer to evolve, which in turn affects resonant absorption. The Kelvin-Helmholtz instability develops where the velocity shear is the strongest, at the edges of the loop perpendicular to the direction of motion. This is important for our study because the growth of such instabilities drains energy from the transverse oscillation, thus leading to increased damping. The effects of cooling appear negligible when looking at the entire evolution even in the high amplitude case, aside from its effect on the period, which is increased due to the lower inertia at the loop apex. With higher density runs, resulting in as small a cooling time as 100 s, the high amplitude run still shows strong damping. \\
The caveats of our study are the lack of a realistic solar atmosphere and the lack of thermal conduction, without which the present hydrodynamic evolution may not be proper (see, e.g. \citealt{1982ApJ...255..783M}). Furthermore, a parametric survey of initial plasma properties, such as temperature, would be insightful. It has been shown \citep{2005A&A...437..311B,2010ApJ...717..163B} that 
during the so-called radiative cooling phase, the losses from the transition region lead to enhanced energy loss from the corona, in the form of an enthalpy flux. This leads to an enhanced mass loss and could enhance the effects of cooling on the oscillations. In addition, thermal conduction could cool the loops even faster. However, the cooling time in the high density runs is short enough to allow for an appreciation of the effects of a higher energy loss. The presence of damping in the high amplitude runs even with fast energy loss indicates that is unlikely that cooling could explain alone the observed, flare related undamped oscillations of coronal loops. These results have implications in the tool of coronal seismology: since the effects of loop cooling with the usually observed cooling times (in our case with $t_{\mathrm{cool}} \approx 800$ s) on the oscillations are negligible, it can also be applied for observations of flare related coronal loop oscillations which show similar cooling behaviour. 
       
\begin{acknowledgements} N.M. acknowledges the Fund for Scientific Research-Flanders (FWO-Vlaanderen). T.V.D. was supported by an Odysseus grant, the Belspo IAP P7/08 CHARM network and the GOA-2015-014 (KU Leuven). A.M. and N.M. was supported by a grant of the Romanian National Authority of Scientific Research, Program for research – Space Technology and Advanced Research – STAR, project number 72/29.11.2013. The software used in this work was developed in part by the DOE NNSA ASC- and DOE Office of Science ASCR-supported Flash Center for Computational
Science at the University of Chicago. Visualization was done with the help of VisIt software \citep{HPV:VisIt}. \end{acknowledgements}

\bibliographystyle{bibtex/aa} 
\bibliography{bibtex/A&Abiblio}{} 

\end{document}